\documentclass[sigconf]{acmart}

\usepackage{amscd,amsfonts,amsbsy,rotating}
\usepackage{balance}
\usepackage{graphicx}
\usepackage{epsfig,epstopdf}
\usepackage{subfigure}
\usepackage{multirow}
\usepackage{booktabs}
\usepackage{color,xcolor}
\usepackage{url}
\usepackage{latexsym,bm}
\usepackage{enumitem,balance,mathtools}
\usepackage{wrapfig}
\usepackage{euscript}
\usepackage{algorithm}
\usepackage{algorithmic}
\usepackage{ifpdf}
\usepackage{diagbox}
\usepackage{caption}
\usepackage{makecell}
\usepackage{subfigure}
\usepackage{bm}

\newcommand{\bs}{\boldsymbol}
\newcommand{\bc}{\bm{c}}

\newcommand{\br}{\bs{r}}
\newcommand{\bbm}{\bs{m}}

\newcommand{\bu}{\bs{u}}
\newcommand{\bv}{\bs{v}}

\newcommand{\bTheta}{\bs{\Theta}}
\newcommand{\bPhi}{\bs{\Phi}}

\theoremstyle{plain}

\newtheorem*{definition*}{Definition}

\DeclareMathOperator*{\argmin}{arg\,min}

\usepackage{array}
\newcolumntype{N}{@{}m{0pt}@{}}

\newcommand{\minisection}[1]{\vspace{5pt}\noindent\textbf{#1.}}

\AtBeginDocument{%
  \providecommand\BibTeX{{%
    \normalfont B\kern-0.5em{\scshape i\kern-0.25em b}\kern-0.8em\TeX}}}

\acmSubmissionID{fp133}

\begin{document}

\copyrightyear{2019} 
\acmYear{2019} 
\setcopyright{acmcopyright}
\acmConference[SIGIR '19]{Proceedings of the 42nd International ACM SIGIR Conference on Research and Development in Information Retrieval}{July 21--25, 2019}{Paris, France}
\acmBooktitle{Proceedings of the 42nd International ACM SIGIR Conference on Research and Development in Information Retrieval (SIGIR '19), July 21--25, 2019, Paris, France}
\acmPrice{15.00}
\acmDOI{10.1145/3331184.3331230}
\acmISBN{978-1-4503-6172-9/19/07}

\settopmatter{printacmref=true}

\title[Lifelong Sequential User Modeling for User Response Prediction]{Lifelong Sequential Modeling with Personalized Memorization for User Response Prediction}

\author{Kan Ren$^1$, Jiarui Qin$^1$, Yuchen Fang$^1$, Weinan Zhang$^1$, Lei Zheng$^1$, Weijie Bian$^2$, Guorui Zhou$^2$, Jian Xu$^2$, Yong Yu$^1$, Xiaoqiang Zhu$^2$ and Kun Gai$^2$}
\authornote{K. Ren, J. Qin and Y. Fang share the co-first authorship. The work was done when they were working as internship in Alibaba Group.}
\affiliation{
  \institution{$^1$Shanghai Jiao Tong University, $^2$Alibaba Group\\
    \{kren, qinjr, arthur\_fyc, wnzhang\}@apex.sjtu.edu.cn, \{weijie.bwj, guorui.xgr, xiaoqiang.zxq\}@alibaba-inc.com\\~}
}

\renewcommand{\shortauthors}{K. Ren, et al.}

\begin{abstract}
User response prediction, which models the user preference w.r.t. the presented items, plays a key role in online services. With two-decade rapid development, nowadays the cumulated user behavior sequences on mature Internet service platforms have become extremely long since the user's first registration. Each user not only has intrinsic tastes, but also keeps changing her personal interests during lifetime. Hence, it is challenging to handle such \textit{lifelong sequential modeling} for each individual user. Existing methodologies for sequential modeling are only capable of dealing with relatively recent user behaviors, which leaves huge space for modeling long-term especially lifelong sequential patterns to facilitate user modeling. Moreover, one user's behavior may be accounted for various previous behaviors within her whole online activity history, i.e., long-term dependency with multi-scale sequential patterns. In order to tackle these challenges, in this paper, we propose a Hierarchical Periodic Memory Network for lifelong sequential modeling with personalized memorization of sequential patterns for each user. The model also adopts a hierarchical and periodical updating mechanism to capture multi-scale sequential patterns of user interests while supporting the evolving user behavior logs. The experimental results over three large-scale real-world datasets have demonstrated the advantages of our proposed model with significant improvement in user response prediction performance against the state-of-the-arts.
\end{abstract}




\maketitle

\section{Introduction}
Nowadays, accurate prediction of user responses, e.g., clicks or conversions, has become the core part in personalized online systems, such as search engines \cite{dupret2008user}, recommender systems \cite{qu2016product} and computational advertising \cite{he2014practical}.
The goal of user response prediction is to estimate the probability that a user would respond to a specific item or a piece of content provided by the online service.
The estimated probability may guide the subsequent decision making of the service provider, e.g., 
ranking the candidate items according to the predicted click-through rate \cite{qu2016product} or 
performing ad bidding according to the estimated conversion rate \cite{zhang2014optimal}.

\begin{figure}[h]
  \centering
  \includegraphics[width=1\columnwidth]{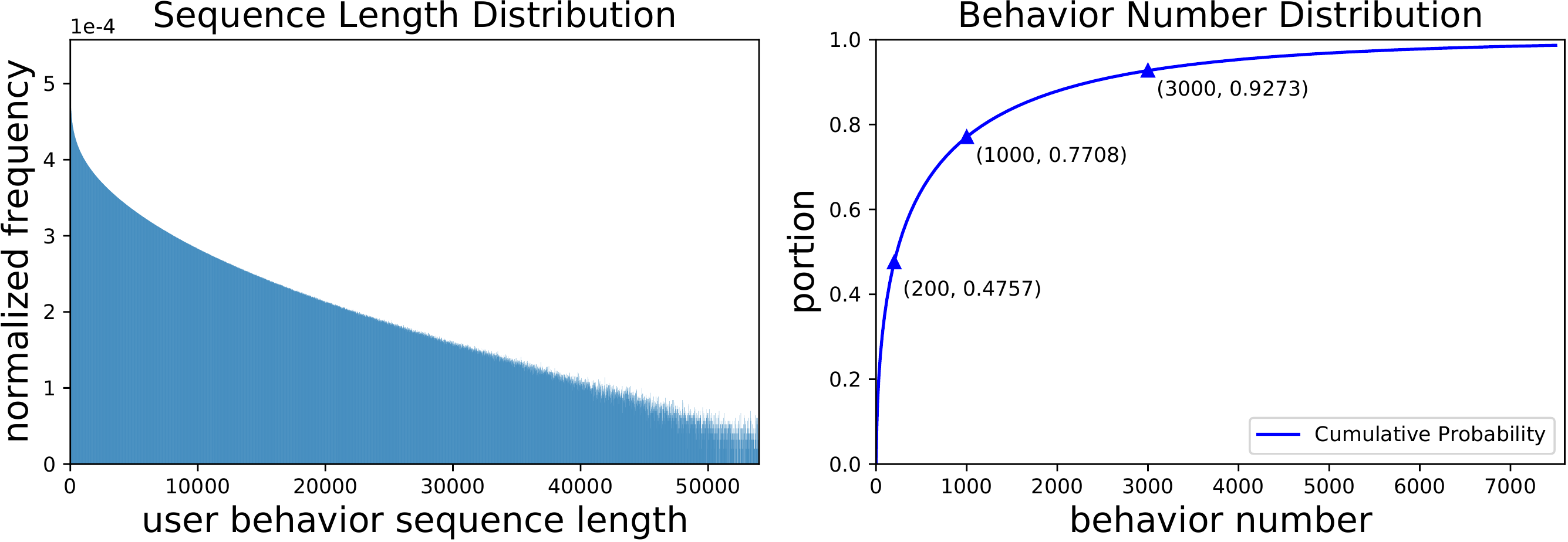}
  \caption{User behavior (click) statistics from Alibaba e-commerce platform during April to September in 2018.
    Left: the distribution of the user sequence lengths;
    Right: the number of user behaviors between add-to-cart event and the final conversion.}
  \label{fig:user-stat}
  \vspace{-10pt}
\end{figure}
One key aspect of user response prediction is user modeling, which profiles each user through learning from her historical behavior data or other side information.
Generally speaking, the user behavior data have three characteristics.
First, the user behaviors not only reflect the intrinsic and multi-facet user interests \cite{jiang2014fema,koren2008factorization}, but also reveal the temporal dynamics of user tastes \cite{koren2009collaborative}.
Second, as is shown in Figure~\ref{fig:user-stat}, the length of behavior sequences vary for different users because of diverse activeness or registration time.
Third, there exist long-term dependencies in one's behavior history where some behaviors happened early may accounts for the final decision making of the user, as illustrated in the right plot of Figure~\ref{fig:user-stat}.
Moreover, the temporal dependency also shows multi-scale sequential patterns, i.e., various temporal behavior dependencies, of different users.

With two-decade of rapid development of Internet service platforms,
there have been abundant user behavior sequences cumulated in online platforms. Many works have been proposed for user modeling \cite{rendle2010factorizing,zhou2018deepa}, especially with sequential modeling \cite{hidasi2017recurrent,zhou2018deepb}.
Some of the existing methods for user modeling aggregate the historical user behaviors for the subsequent preference prediction \cite{koren2009matrix,koren2008factorization}.
However, they ignore temporal dynamics of user behaviors \cite{koren2009collaborative}.
Sequential modeling for user response prediction is to conduct a dynamic user profiling with sequential pattern mining.
Some other works \cite{hidasi2017recurrent,zhou2018deepb} aim to deal with temporal dynamics with sequential pattern mining.
Nevertheless, these sequential models focus only on short-term sequences, e.g., several latest behaviors of the user \cite{zhou2018deepb} or the behavior sequence within recent period of time \cite{hidasi2017recurrent} but abandon previous user behaviors.

Considering the situation of recommending items in the manual way.
Human may first take one's intrinsic tastes into consideration \cite{zhang2018next} and then consider her multi-facet interests \cite{koren2008factorization,jiang2014fema}, e.g., various preferences over different item categories.
Moreover, it is natural to combine one's long-term \cite{ying2018sequential} and recent experience \cite{hidasi2015session} so as to comprehensively recommend items.

In order to tackle these challenges, also to overcome the shortcomings of the related works, we formulate the \textit{lifelong sequential modeling} framework and propose a novel Hierarchical Periodic Memory Network (HPMN) to maintain user-specific behavior memories to solve it.
Specifically, we build a personalized memorization for each user, which remembers both intrinsic user tastes and multi-facet user interests with the learned while compressed memory.
Then the model maintains hierarchical memories to retain long-term knowledge for user behaviors.
The HPMN model also updates memorization from newly coming user behaviors with different periods at different layers so as to capture multi-scale sequential patterns during her lifetime.
The extensive experiments over three large-scale real-world datasets show significant improvements of our proposed model against several strong baselines including state-of-the-art.

This paper has three main contributions listed as follows.
\begin{itemize}[leftmargin=5mm]
  \item To the best of our knowledge, it is the first work to propose the lifelong sequential modeling framework, which conducts a unified, comprehensive and personalized user profiling,
  for user response prediction with extremely long user behavior sequence data.
  \item In lifelong sequential modeling framework, we propose a memory network with incremental updating mechanism to learn from the retained knowledge of user lifelong data and the evolving user behavior sequences.
  \item We further design a hierarchical architecture with multiple update periods to effectively mine and utilize the multi-scale sequential patterns in users' lifelong behavior sequences. 
\end{itemize}

The rest of our paper is organized as below.
Section~\ref{sec:related-work} presents a comprehensive survey of user response prediction works.
Section~\ref{sec:method} introduces the motivation and model design of our methodology in detail.
The experimental setups with the corresponding results are illustrated in Section~\ref{sec:exp}.
We finally conclude this paper and discuss the future work in Section~\ref{sec:conclusion}.

\section{Related Works}\label{sec:related-work}
\subsection{User Response Prediction}
User response prediction is to model the interest of the user on the content from the provider and estimate the probability of the corresponding user event \cite{ren2018bid}, e.g., clicks and conversions.
It has become a crucial part of the online services, such as search engines \cite{dupret2008user}, recommender systems \cite{qu2016product,guo2017deepfm} and online advertising \cite{graepel2010web,zhou2018deepa,he2014practical}.
Typically, user response prediction is formulated as a binary classification problem with user response likelihood as the training objective \cite{richardson2007predicting,graepel2010web,agarwal2010estimating,oentaryo2014predicting}.

From the view of methodology, linear models such as logistic regression \cite{lee2012estimating,gai2017learning} and non-linear models such as tree-based models \cite{he2014practical} and factorization machines \cite{menon2011response,oentaryo2014predicting} have been well studied.
Recently, neural network models \cite{qu2016product,zhou2018deepa} have attracted huge attention.

\subsection{Sequential User Modeling}
User modeling, i.e. to capture the latent interests of the user and derive the adaptive representation for each user, is the key component in user response prediction \cite{zhou2018deepa,zheng2017joint}.
The researchers have proposed many methodologies ranging from latent factor methods \cite{koren2009matrix,rendle2010factorization} to deep representation learning methods \cite{qu2016product,zhou2018deepa}.
These models aggregate all historical behaviors as a whole while ignoring the temporal and drifting user interests.

Nowadays, sequential user modeling has drawn great attention since the sequences of user behaviors have rich information for the user interests, especially with drifting trends.
It has been a research hotspot for sequential modeling in online systems \cite{zhou2018deepb,ren2018learning,villatel2018recurrent}.
From the perspective of modeling, there are three categories for sequential user modeling.
The first is from the view of temporal matrix factorization \cite{koren2009collaborative} with the consideration of drifting user preferences but it heuristically made some assumptions about the behavior patterns.
The second stream is based on Markov-chain methodology \cite{rendle2010factorizing,he2016fusing,he2016vista} which implicitly models the user state dynamics and derive the outcome behaviors.
The third school is based on deep neural network for its stronger capacity of feature extraction, such as recurrent neural network (RNN) \cite{hidasi2015session,hidasi2017recurrent,wu2017recurrent,jing2017neural,liu2016context,beutel2018latent,villatel2018recurrent} and convolutional neural network (CNN) regarding the behavior history as an image \cite{tang2018personalized,kang2018self}.

However, these methods mainly focus on short-term user modeling which has been constrained in the most recent behaviors.
\citet{zhang2018next} additionally utilized a static user representation for user intrinsic interests along with short-term intent representation. \citet{ying2018sequential} proposed a hierarchical attentional method over a list of user behavior features for modeling long-term interests.
But they can only capture simple sequential patterns lacking of considering long-term and multi-scale behavior dependencies.
Moreover, few of the existing works consider modeling lifelong user behavior history thus cannot properly establish a comprehensive user profiling.

\subsection{Memory-augmented Networks}
Memory-augmented networks \cite{hochreiter1997long,weston2015memory,sukhbaatar2015end,kumar2016ask,graves2014neural} have been proposed in natural language processing (NLP) tasks for explicitly remembering the extracted knowledge by maintaining that in an external memory component.
Several works \cite{Ebesu:2018:CMN:3209978.3209991,chen2018sequential,huang2018improving,wang2018neural} utilize memory network for recommendation tasks.
However, these methods directly use the structure of memory network from NLP tasks, which does not consider practical issues in user response prediction.
Specifically, they fail to consider multi-scale knowledge memorization or long-term dependencies.
There is one work of recurrent model with multi-scale pattern mining \cite{chung2016hierarchical} in the NLP field.
The essential difference is that their model was designed for natural language sentence modeling with fixed length, while our model supports lifelong sequential modeling through the maintained user memory and additionally consider long-term dependencies within user behavior sequences with extremely large length.

\section{Methodology}\label{sec:method}
In this part, with discussions about the notations and preliminaries of user response prediction,
we make a definition of lifelong sequential modeling and discuss some characteristics of it.
Then we present the overall architecture of lifelong sequential modeling including the data flow with Hierarchical Periodic Memory Network (HPMN).
The notations have been summarized in Table~\ref{tab:notation}.

\subsection{Preliminaries}
The data in the online system are formulated as a set of triples $\{ (u,v,y) \}$ each of which includes the user $u \in \mathcal{U}$, item $v \in \mathcal{V}$ and the corresponding label of user behavior indicator
\begin{equation}
y = \left\{
  \begin{array}{rcl}
  1, & & u~ \text{has interacted with} ~v; \\
  0, & & \text{otherwise.} \\
  \end{array}
\right.
\end{equation}
Without loss of generality, we take click as the user behavior and the goal is to estimate click-through rate\footnote{In this paper, we focus on the CTR estimation, while the estimation of other responses can be done by following the same tokens.} (CTR) of user $u$ on item $v$ at the given time.
It approaches CTR prediction through a learned function $f_{\bTheta}(\cdot)$ with parameter $\bTheta$.
There are three parts of raw features $(\bu, \bv, \bc)$.
Here $\bv$ is the feature vector of the target item $v$ including the item ID and some side information and $\bc$ is the context feature of the prediction request such as web page URL.
User side feature $\bu = (\bar{\bu}, \{\bv_i\}_{i=1}^{T})$ contains some side information $\bar{\bu}$ and a sequence of user interacted (i.e., \textit{clicked}) items of user $u$.
Note that, the historical sequence length $T$ varies among different users.

The goal of sequential user modeling is to learn a function $g_{\bPhi}(\cdot)$ with parameter $\bPhi$ for conducting a comprehensive representation for user $u$
\begin{equation}\label{eq:user-model}
  \br = g(\{\bv_i\}_{i=1}^{s}; {\bPhi})
\end{equation}
taking the recent $s$ user behaviors.
Note that, this user modeling can be drifting since the user continues interacting with online systems and generating new behaviors.
Many sequential user modeling works set a fixed value $s < T$ as the maximal length of user behavior sequence, e.g., $s=5$ in \cite{hidasi2017recurrent} for session-based recommendation and $s=50$ in \cite{zhou2018deepb}, to capture recent user interests.

Thus the final task of user response prediction is to estimate the probability $\hat{y}$ of user action i.e. click, over the given item as
\begin{equation}
\hat{y} = \text{Pr}(y|\bu, \bv, \bc) = \mathit{f}(\br, \bv, \bc; \bTheta).
\end{equation}

\begin{table}[t]
  \centering
  \caption{Notations and descriptions}\label{tab:notation}
  \resizebox{\columnwidth}{!}{
    \begin{tabular}{c|l}
      \hline
      Notation & Description. \\
      \hline
      $u, v$ & The target user and the target item. \\
      $y, \hat{y}$ & The true label and the predicted probability of user response. \\
      $\bu, \bv, \bc$ & Feature of user $u$, item $v$ and the context information. \\
      $\bar{\bu}$ & The side information of the user. \\
      $\bv_i$ & Feature of the $i$-th interacted item in user's behavior history. \\
      $\br$ & The inferred sequential representation of the user. \\
      $T, D$ & The total behavior sequence length and the layer number of HPMN. \\
      $i, j$ & The index of sequential behavior and network layer ($i \in [1,T], j \in [1,D]$). \\
      $\bbm^j_i$ & The maintained memory content in the $j$-th layer at the $i$-th time step. \\
      $w^j,t^j$ & The reading weight and the period for the $j$-th memory slot\\
      &  maintained by the $j$-th layer of HPMN. \\
      \hline
    \end{tabular}
  }
\end{table}

\subsection{Lifelong Sequential Modeling}\label{sec:lsm}
Recall that, most existing works on sequential user modeling focus on the recent $s$ behaviors, while sometimes $s \ll T$ for the whole user behavior sequence with length $T$.
To the best of our knowledge, few of them consider lifelong sequential modeling.
We define it as below.
\begin{definition*}{Lifelong Sequential Modeling (LSM)}
in user response prediction is a process of continuous (online) user modeling with sequential pattern mining upon the lifelong user behavior history.
\end{definition*}

There are three characteristics of LSM.
\begin{itemize}[leftmargin=5mm]
  \item LSM supports lifelong memorization of user behavior patterns. It is impossible for the model to maintain the whole behavior history of each user for real-time online inference. Thus it requires highly efficient knowledge preserving of user behavior patterns.
  \item LSM should conduct a comprehensive user modeling of both intrinsic user interests and temporal dynamic user tastes, for future behavior prediction.
  \item LSM also needs continuous adaptation to the up-to-date user behaviors.
\end{itemize}

\begin{figure}[t]
  \centering
  \includegraphics[width=1.0\columnwidth]{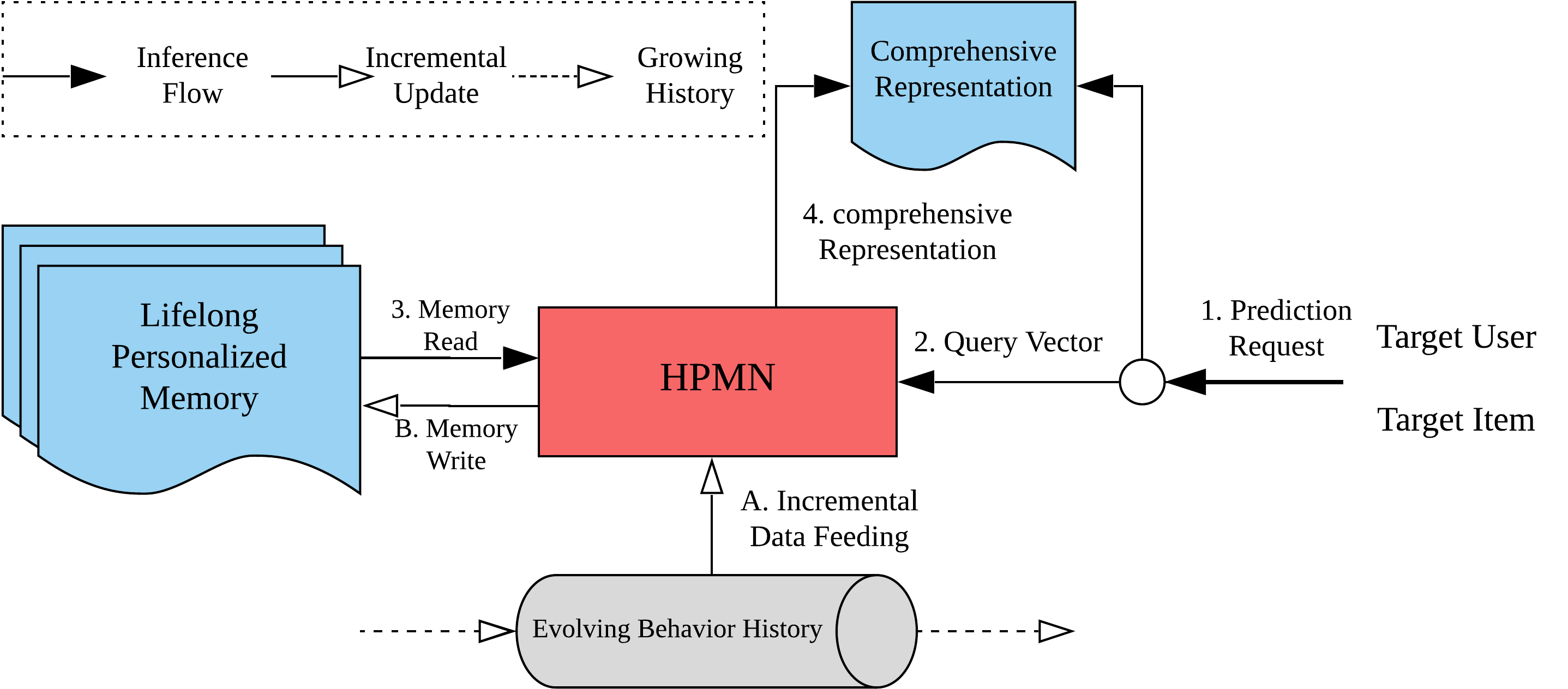}
  \caption{The LSM framework.}
  \label{fig:overall}
\end{figure}

Following the above principles, we propose a LSM framework for the whole evolving user behavior history, as is illustrated in Figure~\ref{fig:overall}.
Within the framework, we conduct a personalized memory with several slots for each user.
This memory will be maintained through an incremental updating mechanism (as Steps A and B in the figure) along with the evolving user behavior history.

As for online inference, when a user sends a visit request, the online service will transmit the request including the information of target user and target item.
Each user request will trigger a query procedure and we use the vector of the target item $\bv$ as the query to obtain the associated user representation according to this specific item in the memory pool.
Then HPMN model will take the query vector to \textit{read} the lifelong maintained personalized memory of that user, to conduct the corresponding user representation, without inference over the whole historical behavior sequence.
After that, the user representation $\br$, item vector $\bv$ and context features $\bc$ will be together considered for the subsequent user response prediction, which will be described in Section~\ref{sec:pred-loss}. \

The details of HPMN will be presented in Section~\ref{sec:hpmn}.

\begin{figure*}[t]
  \centering
  \includegraphics[width=0.95\textwidth]{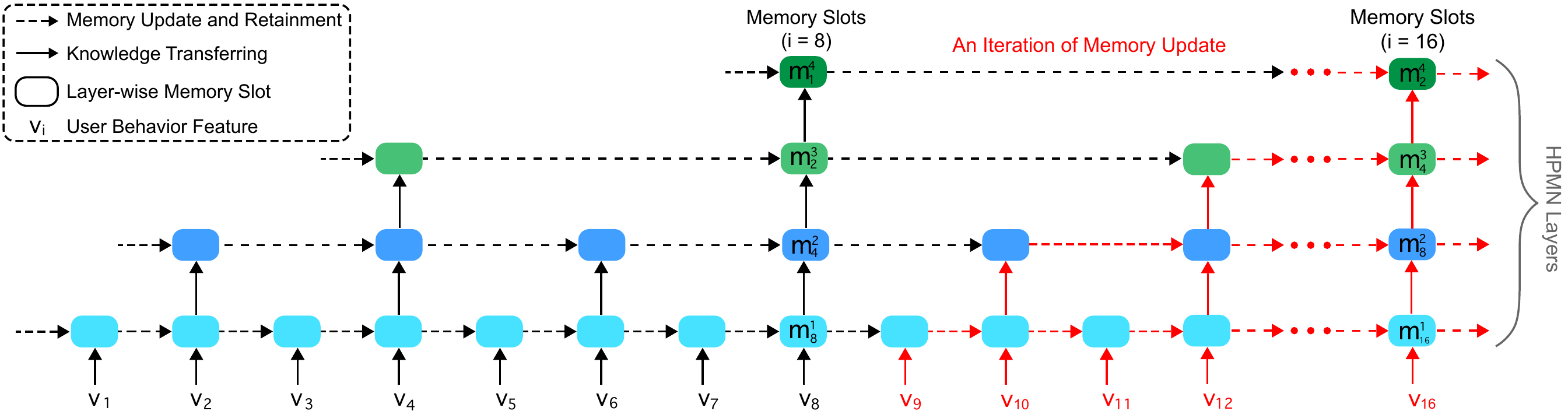}
  \caption{The framework of HPMN model with four layers maintaining user memory in four ($D=4$) memory slots. The update period $t^j$ of $j$-th layer follows an exponential sequence $\{2^{j-1}\}_{j=1}^D$ as an example. The red part means the incremental updating mechanism; the dotted line means the periodic memorization and forgetting.}
  \label{fig:memnet}
\end{figure*}

\subsection{Hierarchical Periodic Memory Network}\label{sec:hpmn}
In this section, we first present the motivations of HPMN model and subsequently discuss the specific architectures.

Generally speaking, we propose HPMN model based on three considerations of the motivation.
\begin{itemize}[leftmargin=5mm]
  \item As is stated above, the main goal of LSM is to capture sequential user patterns hidden in user behavior sequences. Many works \cite{he2016vista,hidasi2015session,zhou2018deepb} have been proposed for sequential pattern mining to improve the subsequent prediction. Thus HPMN model firstly introduces sequential modeling through a recurrent component.
  \item There also exists long-term dependencies among lifelong user behaviors, i.e., the later user decision making may have some relationship to her previous actions. We will show some examples in the experiment of the paper. However, traditional sequential modeling methods either rely on the recent user behaviors, or updates user states too frequently which may result in memorization saturation and knowledge forgetting \cite{sodhani2018training}. Hence we incorporate periodic memory updating mechanism to avoid unexpected knowledge drifting.
  \item The behavior dependencies may span various time distances, e.g., user may show preferences on the specific item at different time along the whole history. So that it requires multi-scale sequential pattern mining. HPMN model deals with this by maintaining hierarchical memory slots with different update periods.
\end{itemize}
Moreover, since the personalized memory stores a comprehensive understanding of each user with multi-facet user preferences, so HPMN model incorporates a regularization of memory covariance to preserve diverse knowledge of user interests.
Besides, for each query, the model reads the user memory through an attentional way which tries to match the target item over the multi-facet user modeling knowledge.

Next we will describe the model details from four aspects.
The memory architecture will be introduced in Section~\ref{sec:mem-arc} followed by the description of periodic yet incremental updating mechanism in Section~\ref{sec:mem-write}.
We introduce the usage of the user memory in Section~\ref{sec:mem-read} and the covariance regularization in Section~\ref{sec:cov-reg}.

\subsubsection{Hierarchical Memory for Sequential Modeling}\label{sec:mem-arc}
As is illustrated in Figure~\ref{fig:overall}, for each user $u$, there is a user-specific memory pool containing $D$ memory slots $\left\{ \bbm^j \right\}_{j=1}^D$ and $\bbm^j \in \mathbb{R}^p$ is a piece of real-value representation of user modeling.
The idea of the external memory has been used in the NLP field \cite{miller2016key,kumar2016ask} for better memorization of the context information embedded in the previously consumed paragraph.
We utilize this external memory pool for capturing the intrinsic user interests with temporal sequential patterns, yet it is also evolving and supports incremental memory update along with the growing behavior sequences.

Generally speaking, HPMN model is a layer-wise memory network which contains $D$ layers, as is shown in Figure~\ref{fig:memnet}.
Each layer maintains the specific memory slot $\bbm^j$.
The output $\bbm^j_i$ of the $j$-th layer at the $i$-th time step (i.e., $i$-th sequential user behavior) will be transmitted not only to the next time step, but also to the next layer at the specific time step.

\subsubsection{Continuous Memory Update}\label{sec:mem-write}
Considering the rapidly growing user-item interactions, it is impossible for the model to scan through the complete historical behavior sequence at each prediction time. That's the reason why almost all the existing methods only consider recent short-term user behaviors.
Thus it is necessary to maintain only the latest memories and implement an incremental update mechanism in real time.
After each user behavior on a item at the $i$-th time step, the memory slot at each layer would be updated as
\begin{equation}
\bbm^j_i =
\left\{
\begin{array}{ccr}
g^j \left(\bbm^{j-1}_i, ~ \bbm^j_{i-1} \right) & { \text{if}~ i ~ \text{mod} ~ t^j ~ = 0 } ~, \\
\bbm^j_{i-1} & {\text{otherwise}} ~,
\end{array}
\right.
\label{eq:mem-write}
\end{equation}
where $j \in \left[ 1, D \right] $ and $t^j$ is the update period of $j$-th layer. 
In Eq.~(\ref{eq:mem-write}), the memory writing in each layer is based on the Gated Recurrent Unit (GRU) \cite{cho2014learning} cell $g^j$ as
\begin{equation}
\begin{aligned}
\bm{z}^j_i &= \sigma (\overline{\bm{W}}^j_z \bbm^{j-1}_i + \overline{\bm{U}}^j_z \bbm^j_{i-1} + \overline{\bm{b}}^j_z) \\
\bm{r}^j_i &= \sigma (\overline{\bm{W}}^j_r \bbm^{j-1}_i + \overline{\bm{U}}_r \bbm^j_{i-1} + \overline{\bm{b}}^j_r) \\
\bbm^j_{i} &= (1 - \bm{z}^j_i) \odot \bbm^j_{i-1} \\
&~~~~~~~~~~~~~~~~ + \bm{z}^j_i \odot \tanh(\overline{\bm{W}}^j_m \bbm^{j-1}_i + \overline{\bm{U}}_m (\bm{r}^j_i \odot \bbm^j_{i-1}) + \overline{\bm{b}}^j_m)  ~.
\end{aligned}
\label{eq:mem-update}
\end{equation}
For each cell in different layers, the parameters $(\overline{\bm{W}}^j, \overline{\bm{U}}^j, \overline{\bm{b}}^j )$ of $g^j$ differs.
Note that, it is a soft-writing operation on the memory slot $\bbm$ since the last operation of $g^j$ function has the ``erase'' vector $\bm{z}^j$ as the same as that in the other memory network literature such as Neural Turing Machine (NTM) \cite{graves2014neural}.
Note that the first layer of memory $\bbm^1_i$ will be updated with the raw feature vector $\bv_i$ of the user interacted item and the memory contents from the last time step $\bbm^1_{i-1}$.

Moreover, the memory update is periodic where each memory $\bbm^j$ at $j$-th memory slot will be updated according to the time step $i$ and the period $t^j$ of each layer.
Here we set the period of each layer $t^j$ as the hyperparameter which is reported in Table~\ref{tab:HPMN-structure} in the experimental setup.
By applying this periodic updating mechanism, the upper layers are updated less frequently to achieve two goals. (i) First it avoids gradient vanishing or explosion, thus being able to model long sequences better; (ii) It then remembers the long-term dependency better than the memory maintained by the lower layer.
The different update behaviors of each layer may capture multi-scale sequential patterns, which is illustrated in Section~\ref{sec:extend-exp}.

The similar idea of clockwork update has been implemented in RNN model \cite{koutnik2014clockwork}.
However, they simply split the parameters in the recurrent cell and update the hidden states separately.
We make two improvements that (i) we connect the network layers through state transferring so as to make layer-wise information transmitting; (ii) we incorporate the external memory component to preserve both intrinsic and multi-scale sequential patterns for lifelong sequential modeling.

\subsubsection{Attentional Memory Reading}\label{sec:mem-read}
Till now, the model has conducted the long-term memorization of the intrinsic and multi-scale temporal dynamics, which may connect the intrinsic properties and the multi-scale patterns of behavior dependency, to the current user response prediction.
Besides, we conduct the attentional memory usage similar to the common memory networks \cite{weston2015memory,sukhbaatar2015end,graves2014neural}.

We calculate the comprehensive user representation $\br$ as
\begin{equation}
\begin{aligned}
\br &= \sum_{j=1}^D w^j \cdot \bbm^j ~.
\end{aligned}
\label{eq:repr-long}
\end{equation}
Here $\bbm^j$ is the maintained memory at the last time step of the long-term sequence, i.e., $i = T$ and $T$ is the final behavior log of the user.
The weight of each memory $w^j$ means the contribution of each memory slot to the final representation $\br$ and it is calculated as
\begin{equation}\label{eq:attn-weight}
w^j = \frac{\exp(e^j)}{\sum_{k=1}^D \exp(e^k)} ~, \text{where } e^j = E(\bbm^j, \bv)
\end{equation}
is an energy model which measures the relevance between the query vector $\bv$ and the long-term memory $\bbm^j$.
Note that the energy function $E$ is a nonlinear multi-layer deep neural network with Rectifier (Relu) activation function $\text{Relu}(x) = \max (0, x)$.
The way we calculate the attention through the energy function $E$ is similar to that in the NLP field \cite{bahdanau2014neural}.

\begin{figure}[t]
  \centering
  \includegraphics[width=0.8\columnwidth]{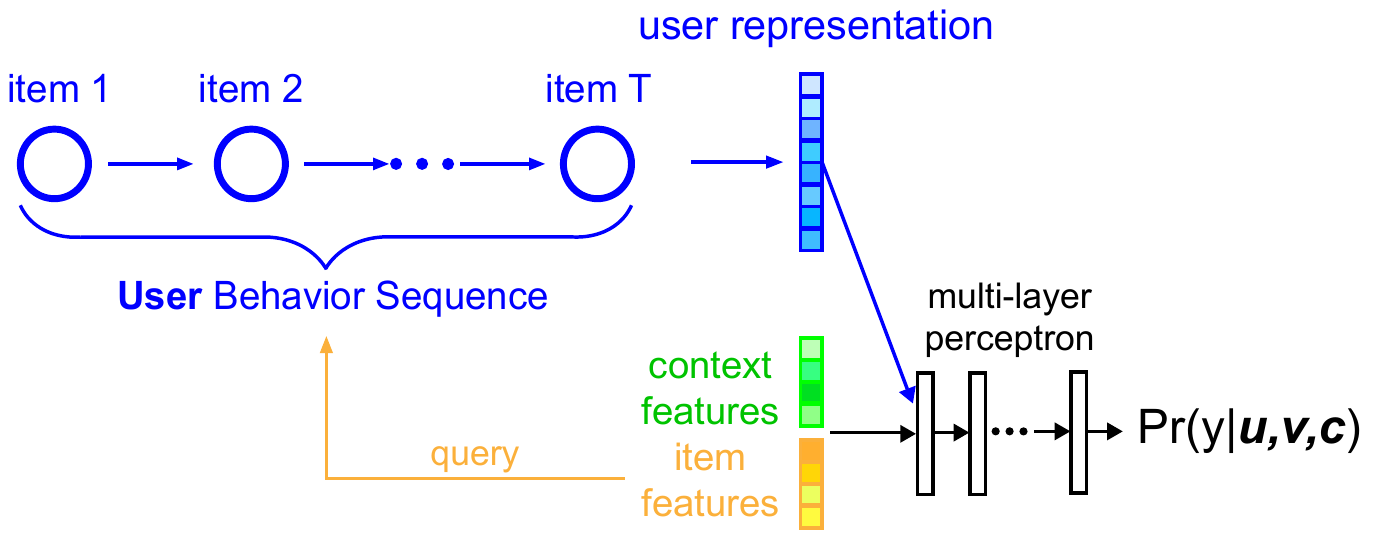}
  \caption{The overall user response prediction.
  }\label{fig:dual}
  \vspace{-10pt}
\end{figure}

\subsubsection{Memory Covariance Regularization}\label{sec:cov-reg}
As is described in the previous sections, the maintained user memory captures long-term sequential patterns with multi-facet user interests.
Recall that our model uses $D$ memory slots with $p$ dimensions to memorize user behavior patterns.
We expect that different memories store knowledge of user interests from different perspectives.
However, unlike the models like NTM \cite{graves2014neural}, HPMN does not utilize attention mechanism to reduce redundancy when updating memory slots.
In order to facilitate memorization utility, we utilize a covariance regularization on memories following \cite{cogswell2016reducing}.

Specifically, we first define $\bm{C}$ as the covariance matrix of the memory contents as
\begin{equation}
  \bm{C} = \frac{1}{p}(\bm{M} - \overline{\bm{M}})(\bm{M} - \overline{\bm{M}})^\top, \text{where } \bm{M}=[\bbm^1,...,\bbm^j,...,\bbm^D]^\top
\end{equation}
is the matrix of memories and $\overline{\bm{M}}$ is the mean matrix with regard to each row of $\bm{M}$ and $p$ is the dimension of each memory slot. Note that $\overline{\bm{M}}$ has the same shape with $\bm{M}$.
After that, we define the loss $\mathcal{L}_c$ to regularize the covariance as
\begin{equation}\label{eq:mem-cov}
\mathcal{L}_c = \frac{1}{2} (\|\bm{C}\|_{F}^2 - \|\text{diag}(\bm{C})\|_2^2)
\end{equation}
where $\|\cdot\|_F$ is the Frobenius norm of matrix.
We need to minimize covariance between different memory slots, which corresponds to penalizing the norm of $\bm{C}$.

\subsection{Prediction Function and Losses}\label{sec:pred-loss}
For each prediction request, we obtain the comprehensive representations $\br$ through querying the personalized memory for the target user by Eqs.~(\ref{eq:user-model}) and (\ref{eq:repr-long}).
The final estimation for the user response probability will be calculated as that in Figure~\ref{fig:dual} as
\begin{equation}\label{eq:dual-func}
  \hat{y} = f(\br, \bv, \bc; \bTheta) ~,
\end{equation}
where $f$ is implemented as a multi-layer deep network with three layers, whose widths are 200, 80 and 1 respectively. The first and second layer use ReLU as activation function while the third layer uses sigmoid function as $\text{Sigmoid}(x)=\frac{1}{1+e^{-x}}$.

As for the loss function, we take an end-to-end training and introduce (i) the widely used cross entropy loss \cite{zhou2018deepa,zhou2018deepb,ren2018bid} $\mathcal{L}_{\text{ce}}$ over the whole dataset with (ii) the covariance regularization $\mathcal{L}_c$ and (iii) the parameter regularization $\mathcal{L}_r$.
We utilize gradient descent for optimization. 
Thus the final loss function is
\begin{equation}
\begin{aligned}
\argmin_{\bTheta, \bPhi} &= \mathcal{L}_{\text{ce}} + \lambda \mathcal{L}_c + \mu \mathcal{L}_r \\
&= -\sum_{k=1}^N \big[ y_k \log \hat{y}_k + (1-y_k) \log (1-\hat{y}_k)\big] \\
&  + \frac{1}{2} \lambda \left( \|\bm{C}\|_{F}^2 - \|\text{diag}(\bm{C})\|_2^2 \right)  + \frac{1}{2} \mu \left( \|\bTheta \|_2^2 + \| \bPhi \|_2^2 \right) ~,
\end{aligned}
\end{equation}
where $\lambda$ and $\mu$ are the weights of the two regularization losses, $\bPhi = \{(\overline{\bm{W}}^j, \overline{\bm{U}}^j, \overline{\bm{b}}^j )\}_{j=1}^D$ is the set of model parameters of HPMN and $N$ is the size of training dataset.

\minisection{Discussions}
We propose the lifelong sequential user modeling with the personalized memory for each user.
The memory are updated periodically to capture long-term yet multi-scale sequential patterns of user behavior.
For user response prediction, the maintained user memory will be queried with the target item to forecast the user preference over that item.

Note that, LSM has some essential differences from the lifelong machine learning (LML) proposed by \cite{chen2016lifelongb}. First, the retained knowledge in LSM is user-specific while LML is model-specific; Second, LSM is conducted for user modeling while LML aims at continuously multi-task learning \cite{chen2016lifelonga}; Finally the user behavior patterns drift in LSM while the data samples and tasks change in LML.

The retained while compressed memory guarantees that the time complexity of our model is acceptable for industrial productions.
The personalized memory will be created from the first registration of the user and maintained by HPMN model as lifelong modeling.
For each prediction, the model only needs to query the maintained memory, rather than inferring over the whole behavior sequence as adopted by the other related works \cite{hidasi2015session,zhou2018deepb}.
Meanwhile, our model has an advantage of sequential behavior modeling to those aggregation-based model, such as traditional latent factor models \cite{koren2009collaborative,koren2008factorization}.
For memory updating, the time complexity is $O(DC)$ where $C$ is the calculation time of the recurrent component.
All the matrix operations can be parallelly executed on GPUs.

The model parameters of HPMN can be updated in a normal way as common methods \cite{qu2016product,zhou2018deepa} where the model is retrained periodically depending on the specific situations.
The number of memory slots $D$ is the hyperparameter and the specific slot number depends on the practical situation.
Along with the lifelong sequential user modeling, the memory of each user is expanded accordingly.
We conduct an experiment about the relations between the number of memory slots and the task performance and discuss in Section~\ref{sec:extend-exp}.
We may follow \cite{sodhani2018training} and expand the memory when the performance drops in some margin.
However, we only need to add one layer with a larger updating period on the top, without retraining all the parameters of HPMN as that in \cite{sodhani2018training}.

\section{Experiments}\label{sec:exp}
In this section, we present the details of the experiment setups and the corresponding results.
We also make some discussions with an extended investigation to illustrate the effectiveness of our model.
Moreover, we have also published our code\footnote{Reproducible code link: https://github.com/alimamarankgroup/HPMN.}.

We start with three research questions (RQs) to lead the experiments and discussions.
\begin{itemize}
  \item [\textbf{RQ1}] Does the incorporation of lifelong behavior sequence contributes to the final user response prediction?
  \item [\textbf{RQ2}] Under the comparable experimental settings, does HPMN achieve the best performance?
  \item [\textbf{RQ3}] What patterns does HPMN capture from user behavior sequences? Does it have the ability to capture long-term, short-term and multi-scale sequential patterns?
\end{itemize}

\subsection{Experimental Setups}
In this part, we present the experiment setups including dataset description, preprocessing method, evaluation metrics, experiment flow and the discussion of the compared settings.

\subsubsection{Datasets}
We evaluate all the compared models over three real-world datasets.
The statistics of the three datasets are shown in Table \ref{tab:dataset-statistics}.
\begin{description}[leftmargin=15pt]
  \item [Amazon] \cite{McAuley2015IRS} is a collection of user browsing logs over e-commerce products with reviews and product metadata from Amazon Inc. We use the subset of Electronic products which contains user behavior logs from May 1999 to July 2014.
  Moreover, we regard all the user reviews as user click behaviors. This processing method has been widely used in the related works \cite{zhou2018deepb, zhou2018deepa}.
  
  \item [Taobao] \cite{zhu2018learning} is a dataset of user behaviors from the commercial platform of Taobao. The dataset contains several types of user behaviors including click, purchase, add-to-cart and item favoring. It is consisted of user behavior sequences from nearly one million users from November 25 to December 3, 2017.
  
  \item [XLong] is sampled from the click logs of more than twenty thousand users on Alibaba e-commerce platform from April to September 2018. It contains relatively longer historical behavior sequences than the other two datasets. Note that there is no public dataset containing such long behavior history of each user for sequential user modeling. We have published this dataset for further research\footnote{Dataset download link: https://tianchi.aliyun.com/dataset/dataDetail?dataId=22482.}.
\end{description}

\noindent\textbf{Dataset Properties}.
These datasets are selected as typical examples in real-world applications.
\textbf{Amazon} dataset covers a very long time range of user behaviors during about fifteen years while some of the users were inactive and generated relatively sparse behaviors during this long time range.
For \textbf{XLong} dataset, each user has a behavior sequence of one thousand clicks that happened in half a year. And modeling such long sequence is a major challenge for lifelong sequential modeling.
As for \textbf{Taobao} dataset, although it only covers nine days' logs, the users in it have generated quite a few behaviors which reflects that the users are quite active.

\noindent\textbf{Dataset Preprocessing}.
To simulate the environment of lifelong sequential modeling, for each dataset, we sort the behaviors of each user by the timestamp to form the lifelong behavior sequence for each user.
Assuming there are $T$ behaviors of user $\bu$, we use this behavior sequence to predict the user response probability at the target item for the $(T+1)$-th behavior. 
Note that 50\% target items at the prediction time in each dataset have been replaced with another item from the non-clicked item set for each user, to build the negative samples.

\noindent\textbf{Training \& Test Splitting}.
We split the training and test dataset according to the timestamp of the prediction behavior.
We set a cut time within the time range covered by the full dataset.
If the prediction behavior of a sequence took place before the cut time, the sequence is put into the training set. Otherwise it would be in the test set. In this way, training set is about 70\% of the whole dataset and test set is about 30\%.

\begin{table}[t]
  \centering
  \caption{The dataset statistics. $T$: length of the whole lifelong sequence (maximal length in the dataset). $s$: length of recent behavior sequence. }\label{tab:dataset-statistics}
  \resizebox{0.55\columnwidth}{!}{
    \begin{tabular}{c|c|c|c}
      \hline
      Dataset & Amazon & Taobao & XLong \\
      \hline\hline
      User \# & 192,403 & 987,994 & 20,000 \\
      \hline
      Item \# & 63,001 & 4,162,024 & 3,269,017 \\
      \hline
      $s$ & 10 & 44 & 232 \\
      $T$ & 100 & 300 & 1,000 \\
      \hline
    \end{tabular}
  }
    \vspace{-10pt}
\end{table}

\subsubsection{Evaluation Metrics}
We use two measurements for the user response prediction task.
The first metric is area under ROC curve (\textbf{AUC}) which assesses the pairwise ranking performance of the classification results between the clicked and non-clicked samples.
The other metric is \textbf{Log-loss} calculated as
\begin{equation}\label{eq:obj-func}\small
\text{Log-loss} = \sum_{k=1}^N \big[- y_k \log \hat{y}_k  - (1-y_k) \log(1 - \hat{y}_k) \big] ~.
\end{equation}
Here $N$ is the number of samples in the test set. Log-loss is to measure the overall likelihood of the whole test data and has been widely used for the classification tasks \cite{ren2016user,qu2016product}.

\subsubsection{Experiment Flow}
Recall that each sample of user behaviors contains at most $T$ interacted items.
As some of our baseline models were proposed to model recent short behavior sequence, thus we first split the recent $s$ user behaviors as the short-term sequential data for baseline model evaluation ($s<T$), as is shown in Table~\ref{tab:dataset-statistics}.
Moreover, for fair comparison, we also conduct the experiments over the whole lifelong sequences with length $T$ for all the baselines.

Note that, all the compared models are fed with the same features including contextual features and side information for fair comparison.

Finally, we conduct the \textbf{significance test} to verify the statistical significance of the performance improvement of our model against the baseline models.
Specifically, we deploy a MannWhitney U test \cite{mason2002areas} under AUC metric, and a t-test \cite{bhattacharya2002median} under Log-loss metric.

\subsubsection{Compared Settings}\label{sec:comp-models}
To show the effectiveness of our method, we compare it with three groups of eight baselines. The first group consists of aggregation-based models, they aggregate the user behaviors for user modeling and response prediction, without considering the sequential patterns.
\begin{itemize}[leftmargin=35pt]
  \item [\textbf{DNN}] is a multi-layer feed-forward deep neural network which has been widely used as the base model in recent works \cite{zhou2018deepa,zhang2016deep,qu2016product}.
  We follow \cite{zhou2018deepa} and use sum pooling operation to integrate all the sequential behavior features concatenating the other features as the user representation.
  \item [\textbf{SVD++}] \cite{koren2008factorization} is a MF-based model that combines the user clicked items and latent factors for response prediction.
\end{itemize}

The second group contains short-term sequential modeling methods including RNN-based models, CNN-based models and a memory network model. For these methods, they either use the behavior data within a session or just truncate the recent behavior sequence to the fixed length.
\begin{itemize}[leftmargin=35pt]
  \item [\textbf{GRU4Rec}] \cite{hidasi2015session} bases on RNN and it is the first work using recurrent cell to model sequential user behaviors. It is originally proposed for session-based recommendation.
  \item [\textbf{Caser}] \cite{tang2018personalized} is a CNN based model, using horizontal and vertical convolutional filters to capture behavior patterns at different scales.
  \item [\textbf{DIEN}] \cite{zhou2018deepb} is a two-layer RNN structure with attention mechanism. It uses the calculated attention values to control the second RNN layer to model drifting user interests.
  \item [\textbf{RUM}] \cite{chen2018sequential} is a memory network model which uses an external memory following the similar architecture in NLP tasks \cite{miller2016key,graves2014neural} to store user's behavior features.
  We implement feature-level RUM as it performed best in the paper \cite{chen2018sequential}.
\end{itemize}

The third group is formed of some long-term sequential modeling methods. However, note that, our HPMN model is the first work on the lifelong sequential modeling for user response prediction.
\begin{itemize}[leftmargin=35pt]
  \item [\textbf{LSTM}] \cite{hochreiter1997long} is the first model to do long-term sequential modeling whose memory capacity is limited.
  \item [\textbf{SHAN}] \cite{ying2018sequential} is a hierarchical attention network. It uses two attention layers to handle user's long- and short-term sequences, respectively. However, this model does not capture sequential patterns.
  \item [\textbf{HPMN}] is our proposed model described in Section ~\ref{sec:method}.
\end{itemize}

We first evaluate the models in the second group over the short length data as they were proposed for short-term sequential modeling. Then we test all the models over the whole length data comparing to our proposed model.

Some state-based user models \cite{rendle2010factorizing,he2016fusing} have been compared in \cite{tang2018personalized} thus we just compare with state-of-the-art \cite{tang2018personalized}.
We omit comparison to the other memory-based models \cite{ebesu2018collaborative,huang2018improving} since they are not aiming at sequential user modeling.

For online inference, all of the baselines except memory models, i.e., RUM and HPMN, need to load the whole user behavior sequence to further conduct user modeling for response prediction, while the memory-based models only need to read the user's personalized memory contents for the subsequent prediction.
Thus the space utility is more efficient of memory-based model considering online sequential modeling.

The difference between our model and the other memory network model, i.e., RUM, is two-fold.
(i) RUM implements the memory architecture following \cite{miller2016key} in NLP tasks, which may not be appropriate for user response prediction, since the user generated data are quite different to language sentences.
And the experiment results in the below section also reflect this.
(ii) Our model utilizes periodic updated memories through hierarchical network to capture multi-scale sequential patterns while RUM has no consideration of that.

\subsubsection{Hyperparameters}\label{sec:hyperparameters}
There are two sets of hyperparameters.
The first set is training hyperparameters, including learning rate and regularization weight. We consider learning rate from $\{1 \times 10^{-4}, 5 \times 10^{-3}, 1 \times 10^{-3}\}$ and regularization weight $\lambda$ and $\mu$ from $\{1 \times 10^{-3}, 1 \times 10^{-4}, 1 \times 10^{-5}\}$. Batch size is fixed on 128 for all the models.
The hyperparameters of each model are tuned and the best performances have been reported below.
The second group is the structure hyperparameters of HPMN model, including size of each memory slot and the update periods $t^j$ of the $j$-th layer which are shown in Table~\ref{tab:HPMN-structure}.
The reported update periods are listed from the first (lowest) layer to the last (highest).

\begin{table}[h]
    \scriptsize
  \centering
  \caption{The HPMN structures on different datasets.}\label{tab:HPMN-structure}
  \resizebox{0.6\columnwidth}{!}{
    \begin{tabular}{c|c|c}
      \hline
      Dataset & Mem. Size & Update Periods\\
      \hline
      Amazon & 32 & 3 layers: 1, 2, 4 \\
      \hline
      Taobao & 32 & 4 layers: 1, 2, 4, 12 \\
      \hline
      XLong & 32 & 6 layers: 1, 2, 4, 8, 16, 32 \\
      \hline
      
    \end{tabular}
  }
  \vspace{-10pt}
\end{table}

\begin{table}[h]
  \centering
  \caption{Performance Comparison. (* indicates p-value < $10^{-6}$ in the significance test. $\uparrow$ and $\downarrow$ indicates the \textit{performance} over lifelong sequences (with length $T$) is better or worse than the same model over short sequences (with length $s$).
  AUC: the higher, the better; Log-loss: the lower, the better.
  The second best performance of each metric is underlined.)}\label{tab:perf-table}
  \resizebox{1.0\columnwidth}{!}{
    \begin{tabular}{c|c|c|lll|lll}
      \hline
      \multirow{2}{*}{Model~Group} & \multirow{2}{*}{Model} & \multirow{2}{*}{Len.} & \multicolumn{3}{c|}{AUC} & \multicolumn{3}{c}{Log-loss}\\
      & & & Amazon & Taobao & XLong & Amazon & Taobao & XLong\\
      \hline
      \multirow{4}{*}{Group 2} &
      GRU4Rec & $s$ & 0.7669 & 0.8431 & 0.8716 & 0.5650 & 0.4867 & 0.4583\\
      & Caser & $s$ & 0.7509 & 0.8260 & 0.8467 & 0.5795 & 0.5094 & 0.4955\\
      & DIEN & $s$ & 0.7725 & 0.8914 & \underline{0.8725} & 0.5604 & 0.4184 & \underline{0.4515}\\
      & RUM & $s$ & 0.7434 & 0.8327 & 0.8512 & 0.5819 & 0.5400 & 0.4931\\
      \hline
      \hline
      \multirow{2}{*}{Group 1} &
      DNN & $T$ & 0.7546 & 0.7460 & 0.8152 & 0.6869 & 0.5681 & 0.5365\\
      & SVD++ & $T$ & 0.7155 & 0.8371 & 0.8008 & 0.6216 & 0.8371 & 1.7054\\
      \hline
      \multirow{4}{*}{Group 2} &
      GRU4Rec & $T$ & 0.7760 $\uparrow$ & 0.8471 $\uparrow$ & 0.8702 $\downarrow$ & 0.5569 $\uparrow$ & 0.4827 $\uparrow$ & 0.4630 $\downarrow$\\
      & Caser & $T$ & 0.7582 $\uparrow$ & 0.8745 $\uparrow$ & 0.8390 $\downarrow$ & 0.5704 $\uparrow$ & 0.4550 $\uparrow$ & 0.5050 $\downarrow$\\
      & DIEN & $T$ & \underline{0.7770} $\uparrow$ & \underline{0.8934} $\uparrow$ & 0.8716 $\downarrow$ & \underline{0.5564} $\uparrow$ & \underline{0.4155} $\uparrow$ & 0.4559 $\downarrow$\\
      & RUM & $T$ & 0.7464 $\uparrow$ & 0.8370 $\uparrow$ & 0.8649 $\uparrow$ & 0.6301 $\downarrow$ & 0.4966 $\uparrow$ & 0.4620 $\uparrow$\\
      \hline
      \multirow{3}{*}{Group 3} &
      LSTM & $T$ & 0.7765 & 0.8681 & 0.8686 & 0.5612 & 0.4603 & 0.4570\\
      & SHAN & $T$ & 0.7763 & 0.8828 & 0.8369 & 0.5595 & 0.4318 & 0.5000\\
      & HPMN & $T$ & \textbf{0.7809}* & \textbf{0.9240}* & \textbf{0.8929}* & \textbf{0.5535}* & \textbf{0.3487}* & \textbf{0.4150}*\\
      \hline
    \end{tabular}
  }
  \vspace{-10pt}
\end{table}

\subsection{Experimental Results and Analysis}
In this section, we present the experiment results in Table~\ref{tab:perf-table} and conduct an analysis from several perspectives.
Recall that the compared models are divided into three groups as mentioned in Sec.~\ref{sec:comp-models}.

\minisection{Comparison between HPMN and baselines}
From Table~\ref{tab:perf-table}, we can tell that HPMN improves the performance significantly against all the baselines and achieves state-of-the-art performance (\textbf{RQ2}).

The aggregation-based models in Group 1, i.e., DNN and SVD++, perform not well as the sequential modeling methods, which indicates that there exist sequential patterns in user behavior data and simply aggregating user behavior features may result in poor performance.

Comparing with the other sequential modeling methods of Group 2, HPMN outperforms all of them regardless of the length of user behavior sequences.
Since GRU4Rec was proposed for short-term session-based recommendation, thus it has the same issue as LSTM which may lose some knowledge of the long-term behavior dependency.
Though the attention mechanism of DIEN improves the performance from GRU4Rec in a large margin, it either ignores the multi-scale user behavior patterns, which will be illustrated from an example in the next section.
Moreover, DIEN model requires to conduct online inference over the whole sequence for prediction, which lacks of practical efficiency considering extremely long, especially lifelong user behavior sequences.
From the results of Caser which uses CNN to extract sequential patterns, we may tell that convolution operation may not be appropriate for sequential user modeling.
As for RUM model, though it utilizes an external memory for user modeling, it fails to capture sequential patterns which results in quite poor performance. Moreover, this proposed model was originally optimizing for other metrics \cite{chen2018sequential}, e.g., precision and recall, thus it may not perform well for user response prediction.

By comparing HPMN with the models in Group 3, i.e., LSTM and SHAN, we find that although both baselines are proposed to deal with long-term user modeling, HPMN has better performance on the very long sequences.
The reason would be that LSTM has limited memory capacity to retain the knowledge, and SHAN has not considered any sequential patterns in the user behaviors.

\minisection{Analysis about Lifelong Sequential Modeling}
Recall that we evaluate all the short-term sequential modeling methods on both short sequence data and lifelong sequence data, as is shown in Table~\ref{tab:perf-table} and we have highlighted the results of the performance gain $\uparrow$ (and drop $\downarrow$) in the table of the latter case compared with the former case.

From the table, we find that almost all the models gain an improvement when modeling on the lifelong user behavior sequences on Amazon and Taobao datasets.
However, on XLong dataset, the performance of GRU4Rec, Caser and DIEN drops, while the memory-based model, i.e., RUM achieve better performance than itself on short sequences.
Note that, our HPMN model performs best.
All the phenomenon reflect that the incorporation of lifelong sequences contributes better user modeling and response prediction (\textbf{RQ1}). Nevertheless, it also requires well designed memory model for lifelong modeling, while our HPMN model achieves satisfying performance on this problem.

\begin{figure}[h]
  \centering
  \includegraphics[width=1.0\columnwidth]{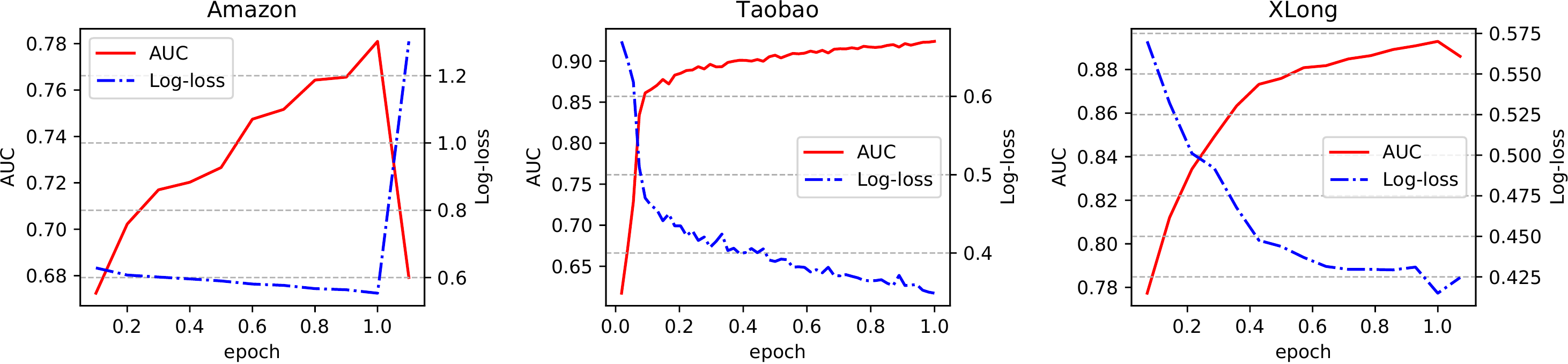}
  \caption{The learning curves on three datasets. Here one epoch means the whole iteration over the training dataset.}\label{fig:lc}
  \vspace{-20pt}
\end{figure}

\minisection{Model Convergence}
We plot the learning curves of HPMN model over the three datasets in Figure~\ref{fig:lc}.
As is shown in the figure, HPMN converges quickly, the Log-loss value on three datasets all drop to the stable convergence after about one iteration over the whole training set.

\begin{figure*}[h]
  \centering
  \includegraphics[width=0.65\textwidth]{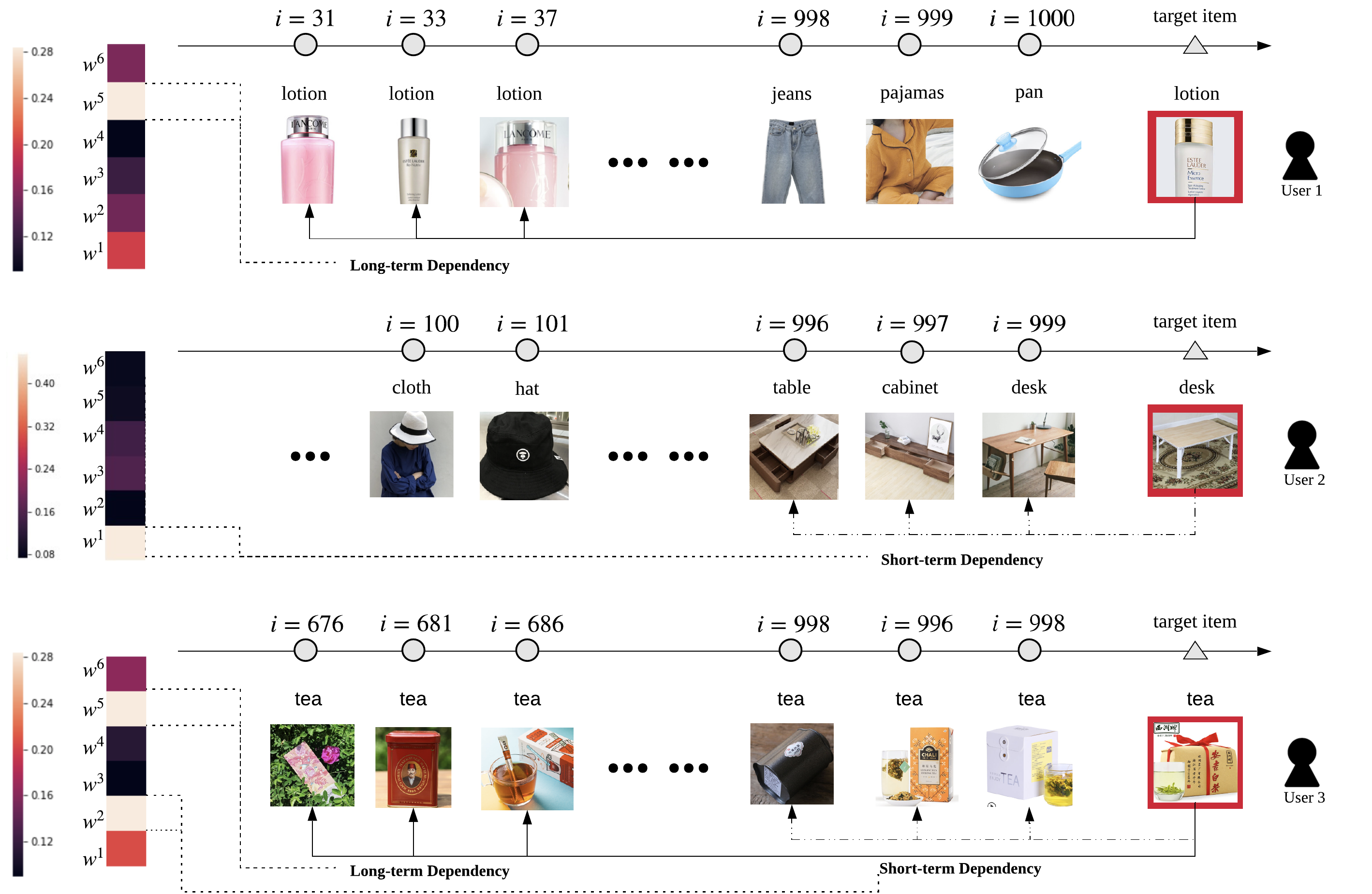}
  \caption{An illustration of long-term, short-term and multi-scale sequential patterns that are captured by HPMN.}\label{fig:atten_seq}
\end{figure*}
\begin{figure}[h]
  \centering
  \includegraphics[width=0.55\columnwidth]{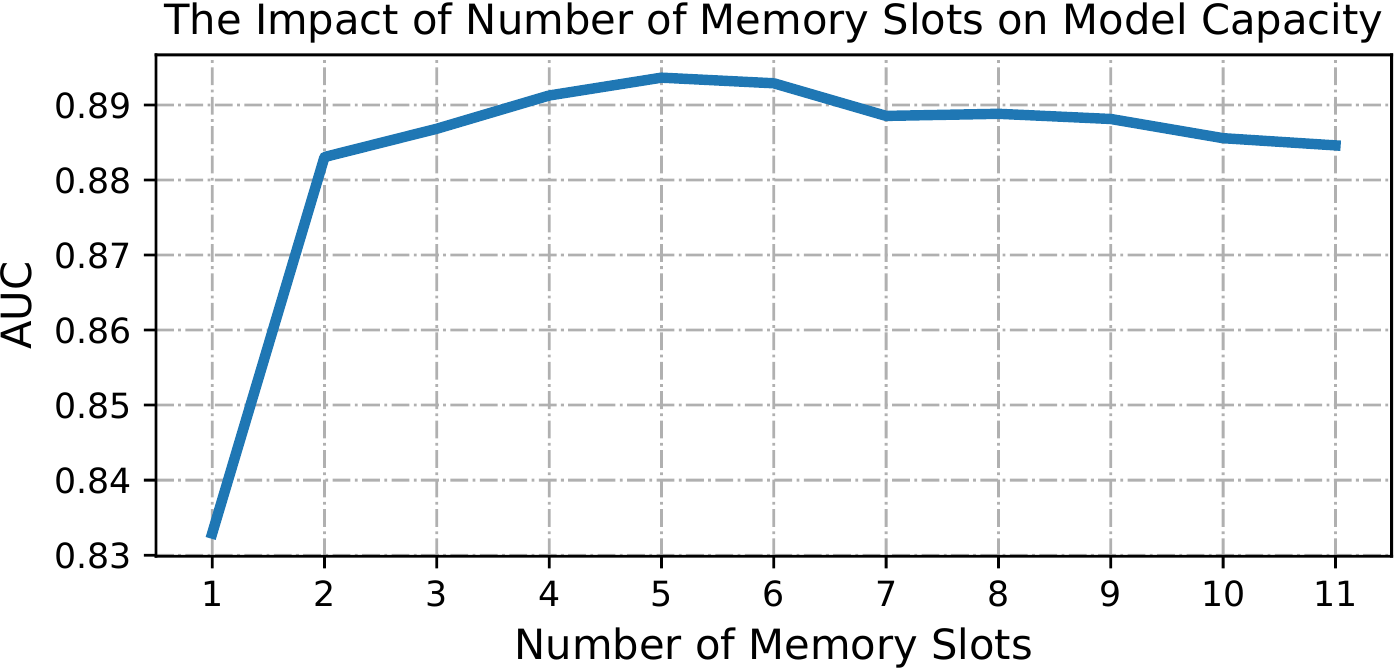}
  \caption{The performance of HPMN with various memory numbers on XLong Dataset. The update period of each $j$-th layer follows exponential sequence $\{2^{j-1}\}_{j=1}^{11}$.}\label{fig:num_of_layer}
  \vspace{-15pt}
\end{figure}

\subsection{Extended Investigation}\label{sec:extend-exp}
In this section, we further investigate the patterns that HPMN captures when dealing with lifelong sequence (\textbf{RQ3}) and the model capacity of memorization.

\minisection{Sequential Patterns with Multi-scale Dependency}
In Figure~\ref{fig:atten_seq}, we plot three real examples of user behavior sequence with length $T=1000$ sampled from XLong dataset.
These three sequences reflect the long-term, short-term and multi-scale sequential patterns captured by HPMN, respectively.

In the first example, the target item is ``lotion'' clicked by the user at the final prediction time. As we find in her behavior history, there are several clicks on lotions at the 31st, 33rd and 37th positions of her behavior sequence, which is far from the tail of her latest behaviors. When HPMN model takes the target item as query to conduct the user representation, from the attention heatmap calculated by HPMN as that in Eq.~(\ref{eq:attn-weight}), we can tell that the fifth layer of HPMN has the maximum attentions, whose update period is relative large. It shows that HPMN captures long-term sequential pattern in the memory maintained by high layers.

In the second example, User 2 at last clicked a desk, and some similar items (table, cabinet) are also clicked in very recent history. However, these kinds of furniture are not clicked in the former part of the sequence. The first memory of HPMN has the maximum attention value which shows that the lower layer is better at modeling short-term pattern for that it updates the memory more frequently to capture user's short-term interests.

As for User 3, the click behavior on the target item has both long-term and short-term dependencies, the similar items are clicked in the recent history and in the former part of her behavior sequence. After inference through HPMN model, the second and fifth layers have higher attention values, for they could capture short-term and long-term dependencies respectively. Thus, this demonstrates that HPMN has the ability to capture multi-scale sequential patterns.

\minisection{Memory Capacity}
In Figure \ref{fig:num_of_layer}, we plot the AUC performance of HPMN with different numbers of memory slots on XLong Dataset. Note that the number of memory slots is equal to the number of HPMN layers.
On one hand, when the number of memory slots 
for each user is less than 5, the prediction performance of the model rises sharply as the memory increases.
This indicates that it requires large memory of sequential patterns for long behavior sequences.
And increasing the memory according to the growth of user behavior sequence helps HPMN to better capture lifelong sequential patterns.
However, when the number of memory slots is larger than 5, the AUC score drops slightly as the memory number increases.
This demonstrates that, on the other hand, the model capacity has some constraints for the specific length of user behavior sequence.
It provides some guides about memory expanding and the principle of enlarging HPMN model for lifelong sequential modeling with evolving user behavior sequence, as is discussed in Section~\ref{sec:pred-loss}.

\vspace{-5pt}
\section{Conclusion}\label{sec:conclusion}
In this paper, we present lifelong sequential modeling for user response prediction.
To achieve this goal, we conduct a framework with a memory network model maintaining the personalized hierarchical memory for each user.
The model updates the corresponding user memory through periodic updating machanism to retain the knowledge of multi-scale sequential patterns.
The user lifelong memory will be attentionally read for the subsequent user response prediction.
The extensive experiments have demonstrated the advantage of lifelong sequential modeling and our model has achieved a significant improvement against strong baselines including state-of-the-art.

In the future, we will adopt our lifelong sequential modeling to improve multi-task user modeling such as prediction of both user clicks and conversions \cite{ma2018entire}.
We also plan to investigate learning for dynamic update period of each layer, to capture more flexible user behavior patterns.

\minisection{Acknowledgments}
The work is sponsored by Alibaba Innovation Research. The corresponding author Weinan Zhang thanks the support of National Natural Science Foundation of China (61702327, 61772333, 61632017) and Shanghai Sailing Program (17YF1428200).

\bibliographystyle{ACM-Reference-Format}
\balance
\bibliography{fp133}

\end{document}